\documentclass[
]{ceurart}

\begin{document}

\copyrightyear{2025}
\copyrightclause{Copyright for this paper by its authors.
  Use permitted under Creative Commons License Attribution 4.0
  International (CC BY 4.0).}

\conference{OPAL 2025: ODRL And Beyond: Practical Applications And Challenges For Policy-Base Access And Usage Control.,
  June 01--02, 2025, Portorož, Slovenia}

\title{Authentication and authorization in Data Spaces: A relationship-based access control approach for policy specification based on ODRL}

\author[1]{Irene Plaza-Ortiz}[%
email=irene.plaza.ortiz@upm.es,
orcid=0009-0008-6382-7482,
]
\fnmark[1]
\author[1]{Andres Munoz-Arcentales}[%
orcid=0000-0003-1554-437X,
email=joseandres.munoz@upm.es,
]
\fnmark[1]
\author[1]{Joaquin Salvachua}[%
orcid=0000-0002-7269-8079,
email=joaquin.salvachua@upm.es,
]
\fnmark[1]
\author[1]{Carlos Aparicio}[%
orcid=0000-0002-6331-3296,
email=carlos.aparicio@upm.es,
]
\fnmark[1]
\author[1]{Gabriel Huecas}[%
orcid=0000-0001-5673-9312,
email=gabriel.huecas@upm.es,
]
\fnmark[1]
\author[1]{Enrique Barra}[%
orcid=0000-0001-9532-8962,
email=enrique.barra@upm.es,
]
\cormark[1]
\fnmark[1]
\address[1]{Departamento de Ingeniería de Sistemas Telemáticos, Information Processing and Telecommunications Center, ETSI Telecomunicaci{\'o}n, Universidad Polit{\'e}cnica de Madrid }

\cortext[1]{Corresponding author.}
\fntext[1]{These authors contributed equally.}

\begin{abstract}
Data has become a crucial resource in the digital economy, fostering initiatives for secure and sovereign data sharing frameworks such as Data Spaces. However, these distributed environments require fine-grained access control mechanisms that balance openness with sovereignty and security. This paper proposes an extension of the Open Digital Rights Language (ODRL) standard, the ODRL Data Spaces (ODS) profile, aimed at supporting authorization —and complementing existing authentication mechanisms— throughout the data lifecycle. Additionally, a policy execution engine is introduced to translate ODRL policies into executable formats, enabling effective enforcement. The approach is validated through a use case involving OpenFGA, demonstrating its applicability to relationship-based access control scenarios. 
\end{abstract}


\maketitle

\section{Introduction}
The European Union’s Data Spaces aim to facilitate secure, decentralized data exchange based on trust and sovereignty. However, effective governance requires enhanced mechanisms for access and usage control \cite{eimear2023european}. ODRL has emerged as a key standard for managing digital rights in these environments \cite{meckler2023web}.

Nevertheless, ODRL existing vocabulary presents limitations when addressing the complex requirements of decentralized and federated environments. This work addresses these challenges by proposing the ODRL Data Spaces (ODS) profile, an extension specifically designed to enhance ODRL's applicability  to cover the life cycle of a data within Data Spaces \cite{akaichi2024interoperable}.

Moreover, as ODRL policies are not inherently executable, this paper introduces a policy execution engine capable of compiling ODRL specifications into executable formats. The integration of these two contributions supports the realization of sovereign and controlled data sharing within dynamic and heterogeneous ecosystems. Additionally, this work explores how OpenFGA, an access control system, complements ODRL by enabling usage control \cite{cimmino2025open}.

While the main focus of this work is on fine-grained authorization, we assume authentication to be handled by existing identity management solutions, such as decentralized identifiers (DIDs) or federated identity providers, which supply trusted identities for policy evaluation.

The paper is structured as follows: Section 2 presents the ODS profile, Section 3 details the execution engine and the use case with OpenFGA, and Section 4 concludes the paper. 

\section{An ODRL Profile for Data Spaces}
\label{sec:sec-odrl-profile}

The proposed ODRL Data Spaces (ODS) profile builds upon existing efforts to enable fine-grained access control in decentralized environments. It extends the Open Access Control (OAC) profile for ODRL \cite{esteves2021odrl}, which itself leverages the Data Privacy Vocabulary (DPV) \cite{w3cDPV}. DPV provides a rich ontology for expressing personal data categories, purposes of processing, and legal bases, facilitating granular policy specification aligned with privacy regulations such as GDPR \cite{DPV}.

However, while OAC and DPV enable detailed modeling of purposes and obligations related to data use, they do not explicitly define the roles, actions, and operational constraints specific to federated Data Spaces. The ODS profile addresses this gap by introducing specialized Party and Action terms tailored to common Data Space scenarios, such as differentiated roles or operational actions like Subscribe, Retention, and Train. These additions support a clearer and more enforceable policy structure in complex data exchange environments. 

Furthermore, the ODS profile is designed to complement existing Data Space architectures, particularly those informed by standards like the International Data Spaces Association (IDSA) Information Model \cite{IDS}. 

Table \ref{tab:my-table-1} includes some of the main terms of the profile, along with a brief description and their alignment with the ODRL vocabulary. 


\begin{table}[]
\centering
\caption{ODRL additional vocabulary}
\label{tab:my-table-1}
\resizebox{\textwidth}{!}{%
\begin{tabular}{ccc}
\textbf{Label} & \textbf{Parent class} & \textbf{Definition}                                  \\ \hline
ODS:Consumer  & odrl:Party            & The Party who acts as the intended user of the data under the Rule. Inherits semantics from odrl:assignee but is specialized for Data Space consumption roles. \\
ODS:Provider   & odrl:Party            & The Party who offers or shares the data asset under the Rule. Specialization of odrl:assigner for Data Spaces. \\
ODS:Broker    & odrl:Party            & The The Party who intermediates data exchanges between Providers and Consumers within the Data Space. \\
ODS:Monitor & odrl:Party            & The Party responsible for overseeing compliance with the Rule, without being directly involved in data usage. \\
ODS:Train      & odrl:Action           & Action to train a machine learning model. \\
ODS:Subscribe    & odrl:Action & Action to subscribe to a dataset, services, or data stream. \\
ODS:Request\_data & odrl:Action  & Action to request specific data from other participants in the data space. \\
ODS:Retention     & odrl:Action  & Action that defines the maximum data retention period before deletion or archiving. \\
ODS:Kill\_job  & odrl:Action           & Action to kill the current executing job. \\ \hline
\end{tabular}%
}
\end{table}

\section{Policy Execution Engine}

While ODRL provides a robust framework for policy specification, its lack of a native execution model limits its applicability in operational environments. To address this, we propose a conceptual policy execution engine that bridges ODRL-based policies with enforcement mechanisms. The approach focuses on compiling ODRL specifications into formats suitable for relationship-based access control systems.

In our prototype, we use OpenFGA —an authorization engine inspired by Google’s Zanzibar— to demonstrate how policies expressed using the ODS profile can be translated into executable rules. OpenFGA is a modern authorization engine designed to be scalable, flexible, and easily integrated into distributed architectures. It supports access policies based on relationships between entities, making it a suitable choice for dynamic and decentralized environments. Moreover, its compatibility with federated identity providers simplifies the authentication of identities, which are mapped to relationships that drive access decisions.

Figure \ref{fig:workflow_openfga} shows the workflow for this use case according to the following process:

\begin{figure}[h]
    \centering
\includegraphics[width=\textwidth]{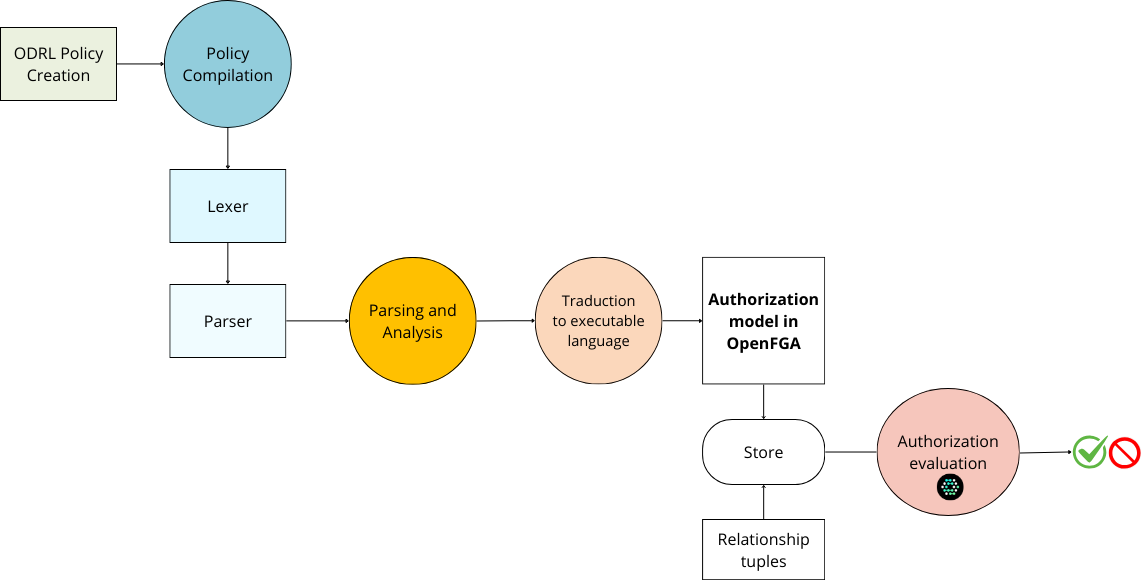}
    \caption{Workflow for a use case with OpenFGA}
    \label{fig:workflow_openfga}
\end{figure}

\begin{enumerate}

    \item ODRL Policy Creation: The first step is to define the access policy using the ODRL language. The policy is set following the previously defined ODS profile. 
    \item Policy Compilation \& Parsing and Analysis: Once the policy is defined in ODRL, it is fed into the policy execution engine, which is responsible for processing and transforming the policy into different executable formats.
    \item Transformation to executable languages: The runtime uses the compiler to translate the ODRL policy into multiple languages. Among these languages, one of the generated formats is compatible with OpenFGA. Although OpenFGA supports both a DSL and a JSON to define authorization models, JSON is the executable format required by OpenFGA. 
    \item Store the Authorization Model: Once the JSON authorization model has been generated, it is stored in the Store. The uploaded model includes the access rules that OpenFGA will use to make authorization decisions.
    \item Store the Permission Tuples: Next, the relationship tuples are written, defining the relationships between users and resources, and stored along with the authorization model in the Store.
    \item Authorization Evaluation: Using the OpenFGA API, permission tuples are queried to assess whether a user has access to a specific resource. The API checks the relationships and rules defined in the authorization model to determine whether access should be granted.
    \item Authorization result: Finally, OpenFGA returns a response indicating whether or not the user has the necessary permission to access the requested resource.

\end{enumerate}

\section{Conclusions and Future work}
This work presents a novel extension of the ODRL framework through the ODS profile, enabling advanced authentication and authorization capabilities tailored to the requirements of Data Spaces. Furthermore, the development of a policy execution engine capable of translating ODRL policies into executable formats provides a practical solution for operationalizing governance mechanisms in decentralized environments. 

Future research will focus on further refining the ODS profile to encompass additional aspects of the data lifecycle \cite{mustafa2024instructions}, integrating compliance with emerging privacy regulations, and optimizing the performance of the policy execution engine in large-scale deployments. One of the key aspects to be incorporated into the profile is the control of data usage, which will be based on the architecture defined in \cite{munoz2019architecture}. 

Another relevant aspect that has not been covered in this article is the study of possible privacy leaks derived from the use of ODRL policies. Although these policies generally do not involve the processing of personal or sensitive data, it is crucial to point out the possibility of privacy leaks through metadata. \cite{pandit2024enhancing}. Special attention will be given to mitigating such privacy risks by carefully managing metadata within policy descriptions, ensuring compliance with the GDPR and similar regulatory frameworks.

By advancing both the expressiveness and enforceability of access control policies, this work contributes to the realization of secure, sovereign, and interoperable Data Spaces. 

\begin{acknowledgments}
The authors would like to acknowledge the support of the FUN4DATE (PID2022-136684OB-C22) project funded by the Spanish Agencia Estatal de Investigacion (AEI) 10.13039/501100011033, and the EUNOMIA Strategic Project: “Solutions for sovereignty, trust and security in data spaces” (C.128.23), result of the collaboration agreement signed between INCIBE and UPM. This initiative is carried out within the framework of the funds of the Recovery, Transformation and Resilience Plan, financed by the European Union (Next Generation).
  \url{https://eunomia.dit.upm.es}.  
\end{acknowledgments}

\section*{Declaration on Generative AI}
  
 During the preparation of this work, the authors used Writefull-AI and DeepL in order to: Grammar and spelling check. After using these tools/services, the authors reviewed and edited the content as needed and takes full responsibility for the publication’s content. 

\bibliography{sample-ceur}

\begin{thebibliography}{11}
\expandafter\ifx\csname natexlab\endcsname\relax\def\natexlab#1{#1}\fi
\providecommand{\url}[1]{\texttt{#1}}
\providecommand{\href}[2]{#2}
\providecommand{\path}[1]{#1}
\providecommand{\DOIprefix}{doi:}
\providecommand{\ArXivprefix}{arXiv:}
\providecommand{\URLprefix}{URL: }
\providecommand{\Pubmedprefix}{pmid:}
\providecommand{\doi}[1]{\href{http://dx.doi.org/#1}{\path{#1}}}
\providecommand{\Pubmed}[1]{\href{pmid:#1}{\path{#1}}}
\providecommand{\bibinfo}[2]{#2}
\ifx\xfnm\relax \def\xfnm[#1]{\unskip,\space#1}\fi
\bibitem[{Eimear et~al.(2023)Eimear, Marco, Alexander, Josep, Brooke, Marina, Monica, Serena, Alessio, Jaime et~al.}]{eimear2023european}
\bibinfo{author}{F.~Eimear}, \bibinfo{author}{M.~Marco}, \bibinfo{author}{K.~Alexander}, \bibinfo{author}{S.~G. Josep}, \bibinfo{author}{T.~Brooke}, \bibinfo{author}{M.~Marina}, \bibinfo{author}{P.~S. Monica}, \bibinfo{author}{S.~Serena}, \bibinfo{author}{T.~Alessio}, \bibinfo{author}{B.~C. Jaime}, et~al., \bibinfo{title}{European Data Spaces-Scientific Insights into Data Sharing and Utilisation at Scale}, \bibinfo{type}{Technical Report}, Joint Research Centre, \bibinfo{year}{2023}.
\bibitem[{Meckler et~al.(2023)Meckler, Dorsch, Henselmann, and Harth}]{meckler2023web}
\bibinfo{author}{S.~Meckler}, \bibinfo{author}{R.~Dorsch}, \bibinfo{author}{D.~Henselmann}, \bibinfo{author}{A.~Harth},
\newblock \bibinfo{title}{The web and linked data as a solid foundation for dataspaces},
\newblock in: \bibinfo{booktitle}{Companion Proceedings of the ACM Web Conference 2023}, \bibinfo{year}{2023}, pp. \bibinfo{pages}{1440--1446}.
\bibitem[{Akaichi et~al.(2024)Akaichi, Slabbinck, Rojas, Van~Gheluwe, Bozzi, Colpaert, Verborgh, and Kirrane}]{akaichi2024interoperable}
\bibinfo{author}{I.~Akaichi}, \bibinfo{author}{W.~Slabbinck}, \bibinfo{author}{J.~A. Rojas}, \bibinfo{author}{C.~Van~Gheluwe}, \bibinfo{author}{G.~Bozzi}, \bibinfo{author}{P.~Colpaert}, \bibinfo{author}{R.~Verborgh}, \bibinfo{author}{S.~Kirrane},
\newblock \bibinfo{title}{Interoperable and continuous usage control enforcement in dataspaces},
\newblock in: \bibinfo{booktitle}{The Second International Workshop on Semantics in Dataspaces, co-located with the Extended Semantic Web Conference}, \bibinfo{year}{2024}.
\bibitem[{Cimmino et~al.(2025)Cimmino, Cano-Benito, and Garc{\'\i}a-Castro}]{cimmino2025open}
\bibinfo{author}{A.~Cimmino}, \bibinfo{author}{J.~Cano-Benito}, \bibinfo{author}{R.~Garc{\'\i}a-Castro},
\newblock \bibinfo{title}{Open digital rights enforcement framework (odre): from descriptive to enforceable policies},
\newblock \bibinfo{journal}{Computers \& Security} \bibinfo{volume}{150} (\bibinfo{year}{2025}) \bibinfo{pages}{104282}.
\bibitem[{Esteves et~al.(2021)Esteves, Pandit, and Rodr{\'\i}guez-Doncel}]{esteves2021odrl}
\bibinfo{author}{B.~Esteves}, \bibinfo{author}{H.~J. Pandit}, \bibinfo{author}{V.~Rodr{\'\i}guez-Doncel},
\newblock \bibinfo{title}{Odrl profile for expressing consent through granular access control policies in solid},
\newblock in: \bibinfo{booktitle}{2021 IEEE European Symposium on Security and Privacy Workshops (EuroS\&PW)}, \bibinfo{organization}{IEEE}, \bibinfo{year}{2021}, pp. \bibinfo{pages}{298--306}.
\bibitem[{W3C(2025)}]{w3cDPV}
\bibinfo{author}{W3C}, \bibinfo{title}{W3c data privacy vocabulary}, \bibinfo{year}{2025}. \URLprefix \url{https://w3c.github.io/dpv/2.1/dpv/}, \bibinfo{note}{accessed on March 25, 2025}.
\bibitem[{J.~Pandit et~al.(2025)J.~Pandit, Esteves, P.~Krog, Ryan, Golpayegani, and Flake}]{DPV}
\bibinfo{author}{H.~J.~Pandit}, \bibinfo{author}{B.~Esteves}, \bibinfo{author}{G.~P.~Krog}, \bibinfo{author}{P.~Ryan}, \bibinfo{author}{D.~Golpayegani}, \bibinfo{author}{J.~Flake},
\newblock \bibinfo{title}{Data privacy vocabulary (dpv) -- version 2.0},
\newblock in: \bibinfo{booktitle}{The Semantic Web -- ISWC 2024}, \bibinfo{publisher}{Springer Nature Switzerland}, \bibinfo{address}{Cham}, \bibinfo{year}{2025}, pp. \bibinfo{pages}{171--193}.
\bibitem[{Mader et~al.(2025)Mader, Pullmann, Petersen, Steffen~Lohmann, IAIS/EIS, and FIT}]{IDS}
\bibinfo{author}{C.~Mader}, \bibinfo{author}{J.~Pullmann}, \bibinfo{author}{N.~Petersen}, \bibinfo{author}{C.~L.-B. Steffen~Lohmann}, \bibinfo{author}{F.~IAIS/EIS}, \bibinfo{author}{F.~FIT}, \bibinfo{title}{W3c international data spaces information model}, \bibinfo{year}{2025}. \URLprefix \url{https://international-data-spaces-association.github.io/InformationModel/docs/index.html#}, \bibinfo{note}{accessed on April 28, 2025}.
\bibitem[{Mustafa et~al.(2024)Mustafa, Nadgeri, Collarana, Arnold, Quix, Lange, and Decker}]{mustafa2024instructions}
\bibinfo{author}{D.~M. Mustafa}, \bibinfo{author}{A.~Nadgeri}, \bibinfo{author}{D.~Collarana}, \bibinfo{author}{B.~T. Arnold}, \bibinfo{author}{C.~Quix}, \bibinfo{author}{C.~Lange}, \bibinfo{author}{S.~Decker},
\newblock \bibinfo{title}{From instructions to odrl usage policies: An ontology guided approach},
\newblock \bibinfo{journal}{Proceedings of the VLDB Endowment. ISSN} \bibinfo{volume}{2150} (\bibinfo{year}{2024}) \bibinfo{pages}{8097}.
\bibitem[{Munoz-Arcentales et~al.(2019)Munoz-Arcentales, L{\'o}pez-Pernas, Pozo, Alonso, Salvach{\'u}a, and Huecas}]{munoz2019architecture}
\bibinfo{author}{A.~Munoz-Arcentales}, \bibinfo{author}{S.~L{\'o}pez-Pernas}, \bibinfo{author}{A.~Pozo}, \bibinfo{author}{{\'A}.~Alonso}, \bibinfo{author}{J.~Salvach{\'u}a}, \bibinfo{author}{G.~Huecas},
\newblock \bibinfo{title}{An architecture for providing data usage and access control in data sharing ecosystems},
\newblock \bibinfo{journal}{Procedia Computer Science} \bibinfo{volume}{160} (\bibinfo{year}{2019}) \bibinfo{pages}{590--597}.
\bibitem[{Pandit and Esteves(2024)}]{pandit2024enhancing}
\bibinfo{author}{H.~J. Pandit}, \bibinfo{author}{B.~Esteves},
\newblock \bibinfo{title}{Enhancing data use ontology (duo) for health-data sharing by extending it with odrl and dpv},
\newblock \bibinfo{journal}{Semantic Web} \bibinfo{volume}{15} (\bibinfo{year}{2024}) \bibinfo{pages}{1473--1498}.

\end{thebibliography}

\end{document}